\documentclass[conference, letterpaper]{IEEEtran}

\usepackage[left = 0.624in, right=0.624in, top = 0.7in, bottom = 0.9in]{geometry}
\usepackage{algorithm, algorithmic}
\usepackage{amssymb,amscd,dsfont,amsmath}
\usepackage{multirow,subfigure}
\usepackage{graphics}
\usepackage{amsmath}
\usepackage{amsfonts}
\usepackage{url}
\usepackage{color}
\usepackage{amsthm}
\usepackage{tikz}

\newcommand{\subparagraph}{}
\usepackage[compact]{titlesec}

\newtheorem{theorem}{Theorem}
\newtheorem{lemma}[theorem]{Lemma}

\newtheorem{definition}{Definition}
\newtheorem{proposition}[theorem]{Proposition}

\newtheorem{corollary}[theorem]{Corollary}

\newfont{\bbb}{msbm10 scaled 500}

\newfont{\bb}{msbm10 scaled 1100}

\newcommand{\bv}{{\bf b}}
\newcommand{\cv}{{\bf c}}

\newcommand{\xv}{{\bf x}}

\newcommand{\zv}{{\bf z}}

\newcommand{\zerov}{{\bf 0}}

\newcommand{\Am}{{\bf A}}
\newcommand{\Bm}{{\bf B}}

\newcommand{\Ec}{{\cal E}}

\newcommand{\Gc}{{\cal G}}
\newcommand{\Hc}{{\cal H}}

\newcommand{\Kc}{{\cal K}}
\newcommand{\Lc}{{\cal L}}
\newcommand{\Mc}{{\cal M}}

\newcommand{\Pc}{{\cal P}}

\newcommand{\Rc}{{\cal R}}
\newcommand{\Sc}{{\cal S}}

\newcommand{\Uc}{{\cal U}}

\newcommand{\Vc}{{\cal V}}
\newcommand{\Xc}{{\cal X}}

\newcommand{\gammav}{\hbox{\boldmath$\gamma$}}

\newcommand{\xiv}{\hbox{\boldmath$\xi$}}

\renewcommand{\det}{{\hbox{det}}}

\DeclareFontFamily{U}{cmfi}{}
\DeclareFontShape{U}{cmfi}{m}{n}{ <-> cmfi10 }{}
\DeclareSymbolFont{CMFI}{U}{cmfi}{m}{n}

\renewcommand{\Am}{\pmb{A}}
\renewcommand{\Bm}{\pmb{B}}

\renewcommand{\bv}{\pmb{b}}
\renewcommand{\cv}{\pmb{c}}

\renewcommand{\xv}{\pmb{x}}

\renewcommand{\zv}{\pmb{z}}

\newcommand{\ind}[1]{\mathds{1}_{\left\lbrace #1 \right\rbrace}}

\begin{document}
	\title{Complexity of URLLC Scheduling and Efficient Approximation Schemes}
	\author{
		\IEEEauthorblockN{Apostolos Destounis and Georgios S. Paschos
			\\}
		\IEEEauthorblockA{Mathematical and Algorithmic Sciences Lab, France Research Center, Huawei Technologies Co. Ltd. \\
			email: firstname.lastname@huawei.com
		}
	}
	\maketitle
	
	\begin{abstract}
		In this paper we address the problem of joint admission control and resource scheduling for \emph{Ultra Reliable Low Latency Communications} (URLLC). We examine two models: (i) the \emph{continuous}, where all allocated resource blocks contribute to the success probability, and  (ii) a \emph{binary}, where only resource blocks with strong signal are ``active" for each user, and user $k$ needs $d_k$  active resource blocks for a successful URLLC transmission. 
		In situations of congestion, we are interested in finding a subset of users that can be   scheduled simultaneously. 
		We show that finding a feasible schedule for at least $m$ URLLC users is NP-complete in  the (easier) binary SNR model, hence also in the continuous. Maximizing the reward obtained from a feasible set of  URLLC users is NP-hard and inapproximable to within ${(\log_2d)^2}/{d}$ of the optimal, where $d\doteq \max_kd_k$. On the other hand, we prove that  checking a candidate set of users for feasibility and finding the corresponding schedule (when feasible) can be done in polynomial time, which we exploit to design  an efficient heuristic algorithm for the general continuous SNR model. We complement our theoretical contributions with a numerical evaluation of our proposed schemes. 
	\end{abstract}

	\section{Introduction}
	
	\subsection{Motivation and Background}
	
	
	A key differentiator of upcoming 5G wireless networks is their ability to provide reliable low latency via the \emph{Ultra Reliable Low Latency Communications} (URLLC) service class \cite{3gppURLLC}. 
	This capability
	is considered as an enabler for  industrial automation \cite{Holfeld2016}, virtual reality applications \cite{Zhang2017},  and control of vehicles \cite{Campolo2017}.
	Such applications require ``live'' wireless connections, where packets must be received within a very short time period since their creation.
	To effectively synchronize industrial machines and avoid car collisions, the URLLC requirement not only ensures that packets arrive in time, but also in a reliable manner, in the sense that the latency deadline may be violated only very rarely (e.g. once every 100k attempts). 
	To achieve the URLLC requirement,  5G wireless networks will employ various intelligent techniques, including interface diversity \cite{Nielsen2018}, multi-path diversity \cite{Kotaba2018}, packet duplication \cite{Rao2018} and short-packet communications \cite{Durisi2016}.   \emph{In this paper we focus on scheduling URLLC short packets.}
	
	
	
	Previous wireless schedulers, designed for high bandwidth applications, assigned the \emph{Resource Blocks} (RBs) opportunistically one-by-one  to the user with the highest ratio of \emph{instantaneous rate} over the \emph{average throughput obtained thus far}. Such a simple and efficient algorithm achieves the optimal performance in that setting \cite{Huang2009}. However,  to  optimally schedule  a URLLC user, a radically different approach must be taken;  the available resource blocks within a \emph{Transmission Time Interval} (TTI) are proactively examined and an allocation is made such that the combination of the allocated blocks allow a URLLC  user to achieve its reliability requirement. When multiple URLLC users are served by the same scheduler, a joint allocation of URLLC transmissions must be found on the available resource blocks such that the requirements of all users are satisfied. Therefore, URLLC scheduling consists in combinatorial allocation of resource blocks, which is  a challenging setting for scheduling. This brings us to the natural complexity question: \emph{are there efficient URLLC schedulers?}
	
	
	A further complication arises when a scheduler must serve a set of URLLC users whose requirements are not simultaneously achievable. In this case, it is possible to reject some of the users, and then schedule the rest. However, identifying the optimal schedulable subset of users is shown in this paper to be an extremely difficult problem,  impossible to resolve exactly under tight timing constraints.
	
	More generally,  this paper highlights a crucial consideration towards a theory of scheduling and admission control for guaranteed latency in wireless networks,  that of complexity, which determines how  feasible it is for a practical wireless system to operate with a given algorithm. The aim of this paper is thus to lay the foundations of understanding  the complexity of URLLC communications. Specifically, our contributions include:
	\begin{itemize}
		\item We  model  URLLC scheduling at two different granularities, \emph{(i)} the standard continuous Signal-to-Noise Ratio (SNR) model, and \emph{(ii)} the binary SNR model, an approximation where  each resource block is classified as active or inactive according to an SNR threshold.
		\item We show that the decision problem: \emph{does there exist a URLLC schedule that satisfies $\geq m$ users within a TTI?}  is NP-hard for both SNR models. The statement is proved by  a reduction from the independent set problem, which allows us to characterize also the inapproximability of the corresponding optimization problem. 
		\item For scheduling in the binary SNR model, however, we prove that given a set of URLLC users which is feasible, a schedule can be found in polynomial time solving a linear program. This remarkable simplification is due to the fact that the constraint matrix of the linear relaxation of our scheduling problem is shown to be \emph{totally unimodular}.
		\item Regarding the admission control in the binary model,  we show that the GREEDY algorithm provides a $1/(d+1)$ approximation to the original problem.
		\item Last,  we propose the \emph{Iterative Thresholding Algorithm} (ITA), which applies the above findings to the continuous SNR model. In our simulations  ITA outperforms the continuous greedy baseline by up to $30\%$.
	\end{itemize}
	
	\subsection{Related Work}
	
	
	The quest for low latency wireless communications has gained significant attention in the literature. 
	Works \cite{She2017, She2018} propose a queueing framework and model the URLLC reliability constraint as the probability that the delay of a packet exceeds a threshold. Their  algorithms are based on converting the delay into throughput by the theory of \emph{effective capacity}. 
	The work \cite{Arnau2018} deals with obtaining bounds on delay violation probabilities for a single-user system, where the transmitter employs multiple antennas and short codewords by using \emph{stochastic network calculus.} 
	The above works assume that packets exceeding deadlines still count towards the system performance. When packets that arrive after the  deadline are dropped, authors in \cite{Destounis2018} model the URLLC problem based on the \emph{timely throughput} approach 
	and focus on meeting  a long term packet delivery rate.
	
	
	Regarding scheduling in systems with multiple resources in the time-frequency domain, which is the focus of the current paper, authors in \cite{Sharma11, Bodas2012} address scheduling in the frequency domain in a binary SNR model under a delay violation requirement. Regarding hard deadlines, authors in \cite{Anand2018_puncturing} examine the problem of maximizing the utility of \emph{enhanced Mobile Broadband} (eMBB) 
	users when URLLC transmissions are being \emph{punctured} in resource blocks in the time-frequency grid of LTE. In addition, the work \cite{Anand2018_ofdma} examines the impact of resource allocation in the frequency domain coupled with \emph{Hybrid Automatic Repeat reQuest} (HARQ) on how can a system support a load of URLLC users under queueing theoretic blocking models. Both these works assume that  URLLC users need a fixed number of resources for successful transmission regardless of the actual realization of the channel 
	in each resource block. 
	Finally,  \cite{Fountoulakis17_scalable} examines the impact of dynamically varying TTI length in order to serve URLLC before their deadline and still give enough utility to the eMBB users, assuming, however that wireless transmissions cannot fail, therefore not accounting for small blocklength transmissions.     
	
	Contrary to the aforementioned works, we examine joint admission control and scheduling of radio resource blocks where we take into account the realization of the wireless channel within each block and the transmission failure probabilities due to short block length transmissions. 
	More importantly, our work is the first to characterize the computational complexity and examine approximation schemes for the problem of joint admission control and scheduling of URLLC users in a time-frequency resource grid with fixed resources.
	
	\section{System Model}\label{sec:system}
	
	We consider a system with  $K$ users, operating in frames.  Each frame consists of $R$ \emph{Resource Blocks} (RBs) in the time - frequency domain.
	 Let  $\gamma_r^k$  denote the user-$k$ SNR  in  RB $r$ within the current frame. Values $\gamma_r^k$ are made known to the scheduler via measurements.
	The goal of the scheduler is to assign RBs to each user to satisfy their latency requirements.
	
	
	The latency requirement of URLLC in the 5G specifications is $1ms$, and equal to the frame length \cite{3gpp5GNumerology}. Therefore, one way to satisfy   the user-$k$ latency requirement   would be to correctly communicate $L_k$ bits within each frame.
	However, due to unavoidable transmission errors, correct reception  can not be ensured in a wireless system at all times. To address this inherent limitation of wireless systems, it is meaningful  to consider a probabilistic \emph{Service-Level Agreement} (SLA) in the following  form:
	
	\begin{definition}[URLLC SLA]
		We say that the URLLC SLA of user $k$ is satisfied in a given  frame if:
		\[
		\text{Pr}\left(L_k \text{ bits correctly received}\right)\geq \theta_k.
		\]
		In 5G specifications, $\theta_k =0.99999$ and $L_k=32${ Bytes} \cite{3gppURLLC}. 
	\end{definition}

	In order to meet the above URLLC SLA, the scheduler assigns a set of RBs to each user using the scheduling variables $x_r^k\in\{0,1\}$, where $x_r^k=1$ denotes that user $k$ is scheduled to transmit in RB $r$. Assuming a user can exploit multiple assigned RBs to jointly encode messages, the frame error probability defined as
	\[p_e^k(\xv)\triangleq 1-\text{Pr}\left(L_k \text{ bits correctly received}\right),\]
	depends on the assigned RBs $\xv=(x_r^k)$: allocating more to user $k$ will decrease its frame error probability. Specifically, 
	an accurate estimate of user-$k$ frame error probability  for a  schedule $\xv$  can be found via the 
	Polyanskiy bound \cite{Polyanskiy10}, \cite{Yang2014_multichannel}:
	\begin{align} \nonumber 
	p_e^k(\xv) &= \\ \label{eq:errorProb}
	& \hspace{-0.3in} Q\left(\frac{n\sum_{r}x^k_{r}\log_2(1 + \gamma_{r}^k) - L_k + 0.5\log_2(n)\sum_{r}x_{r}^k}{\sqrt{n\sum_{r}x^k_{r}V(\gamma_{r}^k)}}\right)
	\end{align}
	where $n$ is the number of channel uses, $Q(.)$ is the error function, and $V(\gamma) = 1 -\frac{1}{(1+\gamma)^2}$ is the  dispersion of the \emph{Additive White Gaussian Noise} channel with SNR $\gamma$ \cite{Polyanskiy10}. 
	Eq. \eqref{eq:errorProb} is significantly more accurate than the Shannon formula when the length of the transmitted packets is short, such as in our  URLLC case.
	
	In our system, the feedback about transmission failures is obtained at the end of the frame.\footnote{Exploiting feedback within frames is avoided in practical systems since it complicates decision making, and the available time between slots is minimal.}  
	Therefore, to satisfy the user-$k$ URLLC SLA, the scheduler must  pro-actively schedule enough RBs to  provide sufficiently  low error probability, taking into account  the user-specific SNRs $\gamma_r^k, r=1,\dots, R$. Assigning  RBs to users is called \emph{URLLC scheduling},  \emph{and the focus of this paper is URLLC scheduling for $K$ users.}
	
	\subsection{Binary SNR Model}
	
\begin{figure}
		\centering 
		\includegraphics[scale = 0.25]{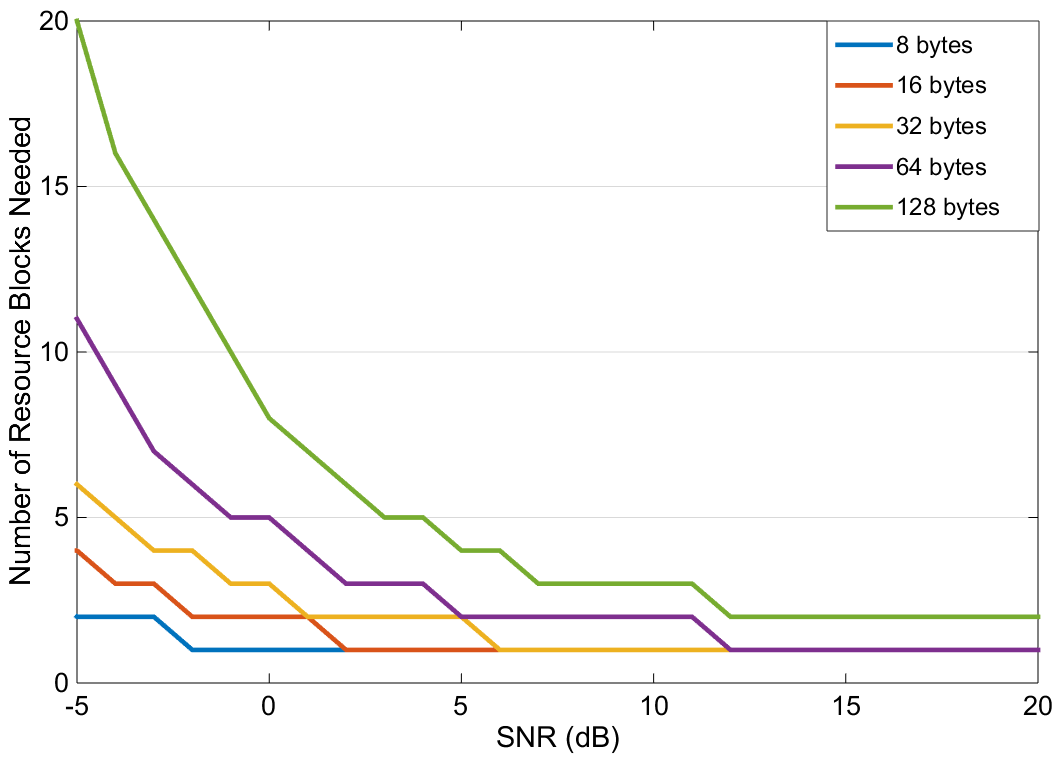}
		\caption{Number of resource blocks needed for $99.999\%$ reliability for different packet sizes in 5G NR. All resource blocks are  assumed to have  the same SNR. 
		} \label{fig:resourceBlocksNeeded}	
	\end{figure}

	To obtain insight into our scheduling problem, we insert numbers in the above formulas from the 5G specifications\footnote{We mention that the number of channel uses are computed based on taking half a 5G subframe as a scheduling unit in the time domain, i.e., $12$ subcarriers and $7$ symbols per resource block; this is an envisioned strategy for supporting low latency traffic in 5G \cite{3gpp5GNumerology}.} and deduce  the number of RBs required to satisfy the URLLC SLA for given $L$ and $\gamma$ (assuming $\gamma_r^k=\gamma, \forall r,k$), shown in Fig.~\ref{fig:resourceBlocksNeeded}. For $32$ Bytes, as few as $3$ RBs with $\gamma > 0dB$ are enough to guarantee the required $99.999\%$ reliability. Many SNR values lead to the same result, and hence the exact value of the SNR may not be crucial. Instead, we will often require that each user is assigned a large enough number (here 3) of  RBs with strong SNR (e.g.~$>0 dB$). 
	
	The above motivates us to study the binary SNR model. We say RB $r$ is  \emph{active} for user $k$ (hence in state  $1$)  if $\gamma_r^k$ is larger than a threshold, and inactive (state $0$) otherwise,  and additionally $d_k$ active RBs  suffice for user $k$  to achieve the required packet error probability. Essentially we have used a user-specific SNR threshold  and treat all values below the threshold as zero. 
	In the following, we  set $\delta_r^k=1$ if $r$ is active for $k$, and 0 otherwise. This model is called the \emph{binary SNR model}, while the previous one will be called continuous.
	
	We show that \emph{(i)} the joint URLLC admission control and scheduling  problem remains complex under the binary SNR model,  \emph{(ii)} the difference between binary and continuous model is small in the LTE/5G specifications, and \emph{(iii)} we may use the results of the binary model to  obtain an efficient algorithm for  the continuous model under any specifications.

	\subsection{URLLC Feasibility and Scheduling}
	
	An essential constraint for scheduling is to allocate each RB to at most one user, written as 
	\begin{equation}\label{cstr:2}
	\sum_{k}x_{r}^k \leq 1,  \quad r=1,\dots,R.
	\end{equation}
	Hereinafter, consider the set of \emph{URLLC schedules}:
	\[
	\mathcal{X}\triangleq \left\{\xv \in \{0,1\}^{K\times R}~\big|~\eqref{cstr:2} \text{ satisfied}\right\}
	.\]
	Next, we are interested in  URLLC schedules $\xv\in\Xc$ that also support the  SLA of a set of users $\Kc$. Specifically, that the SLA of  user $k$ is satisfied in the continuous SNR model if
	\begin{equation}\label{cstr:01}
	p_e^k(\xv) \leq 1 - \theta,  
	\end{equation}
	and in the binary SNR model if
	\begin{equation}\label{cstr:1}
	\sum_{r}x_r^k \geq d_k.
	\end{equation}
	
	\begin{definition}[Feasibility]
		We say that  a URLLC schedule $\xv\in\Xc$ is feasible for users $\Mc$ in the binary (continuous) SNR model  if eq. \eqref{cstr:1} (eq. \eqref{cstr:01}) is satisfied for all $k\in \Mc\subseteq \Kc$. The set of all such feasible schedules is denoted with $\Xc(\Mc)\subseteq \Xc$.
	\end{definition}
	
	Given a set of users $\Kc$, their SLAs, and their SNRs $\gammav$,  an important question regards the URLLC scheduling feasibility: 
	
	\vspace{0.05in}\begin{center}\emph{Q1: is there a  feasible schedule for $\Kc$?}\end{center}
	\vspace{0.05in}
	
	Additionally, if the answer to Q1 is ``yes'' and hence $\Xc(\Kc)$ is non-empty, then we would like to find an $\xv\in \Xc(\Kc)$, e.g., by solving the following feasibility problem:
	
	\vspace{0.1in}
	\noindent\underline{\emph{URLLC Scheduling:}}
	\begin{equation}\label{eq:scheduling}
	\min_{\xv\in \Xc(\Kc)}0
	\end{equation}
	Both Q1 and \eqref{eq:scheduling}  involve searching in a combinatorial space  exponential to $K\times R$, and therefore are possibly complex to address. Unexpectedly, in Section \ref{sec:effsch} we show that under the binary SNR model both questions can be addressed in polynomial time.

	%
	%
	%
	%

	\subsection{URLLC Admission Control}
	
	Next we consider the case that the answer to Q1 is ``no'', i.e., when scheduling all users in $\Kc$ is infeasible. 
	In this case, we are interested in the following admission control question. 
	\begin{center}
		\noindent \emph{Q2: is there a schedule that satisfies at least $m$ users?}
	\end{center}
	\vspace{0.05in}
	We also consider  a more general  approach, where we assign to  users non-negative utilities $w^k, ~k\in\Kc$, which are collected only for users with satisfied  SLAs. We would like to choose the schedule $\xv\in\Xc$ that  collects the maximum total utility, which corresponds to ensuring the URLLC SLA for the most important users. 
	The user-specific utilities can be tweaked according to the application, in order to provide preferential admission of users into the system; for instance, high utility users may correspond to remote controlled vehicles. We introduce the admission variable $z^k\in\{0,1\}$, which takes value 1 if user $k$ will be served and 0 otherwise.
	Then we consider the following optimization:
	
	\vspace{0.1in}
	\noindent\underline{\emph{URLLC Utility Maximization (UUM):}}
	\begin{align}
	& \text{maximize}_{\xv\in\Xc, \zv}
	\sum_{k=1}^Kw^kz^k\label{eq:maxutil}\\
	\text{s.t.} & \sum_{r}x_r^k \geq d_k z^k,\quad k=1,\dots,K \label{eq:cstr1}\\
	& x_r^k \leq \delta_r^k,~~ \quad\quad\quad\forall (r, k) \label{eq:cstr255}\\
	&  x_r^k \in\{0,1\},~~ \forall (r, k) \quad z^k \in \{0,1\},~~ \forall k.\label{eq:cstr3}
	\end{align}
	Note that the linear objective  \eqref{eq:maxutil} drives the solution towards the  $\Mc$-SLA feasible schedule that collects  the highest utility. Constraint \eqref{eq:cstr1} ensures the SLA satisfaction of all selected  users  with $z^k=1$ (it should be replaced with $p_e^k(\xv)+\theta-1 \leq 1-z^k$ for the continuous SNR model). 
	Constraint \eqref{eq:cstr255} restricts schedules on active resource blocks (should be omitted in continuous SNR),  and  \eqref{eq:cstr3} forces the variables to be integers.
	
For the binary model,  \eqref{eq:maxutil}-\eqref{eq:cstr3} is an Integer Linear Program (ILP) of possibly large dimensions. In  Section \ref{sec:complexity} we show that Q2 and the UUM problem are both complex to address exactly. Then in Sections \ref{sec:approx_bin}-\ref{sec:approx_con} we provide approximations for both models that are polynomial-time computable.
	
	%
	%
	%
	%
	%
	
	\section{Complexity of URLLC Admission Control}\label{sec:complexity}
	
	
	\begin{figure*}[h!]
		\centering 
		\subfigure[URLLC graph]{\includegraphics[width=1.82in]{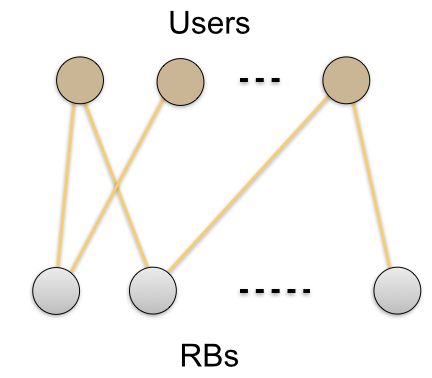}\label{Fig:step1}}
		\quad
		\subfigure[Abstract graph $\Gc$]{\includegraphics[width=2.2in]{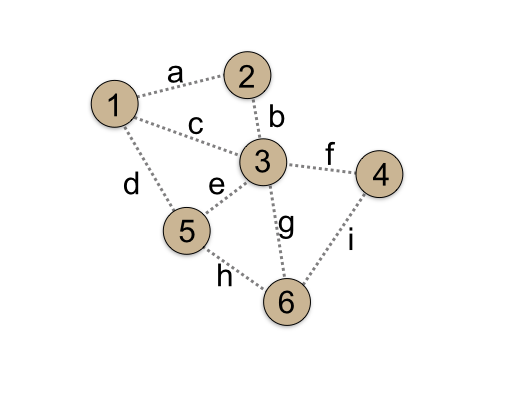}\label{Fig:step2}}
		\quad
		\subfigure[Bipartite graph $\Hc(\Gc)$]{\includegraphics[width=2.5in]{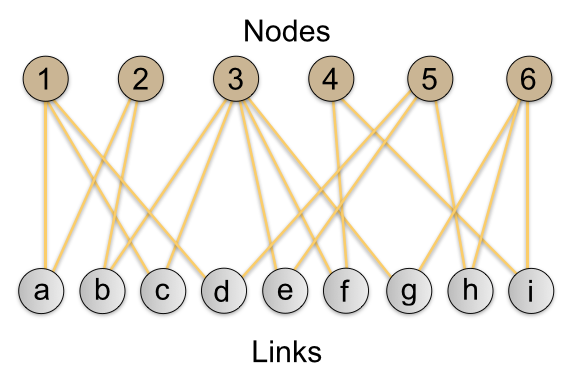}\label{Fig:step3}}
		\caption{(a) Bipartite model of the URLLC scheduling problem. (b) Graph $G$ on which we want to compute a maximum independent set. (c) Bipartite graph modeling the connectivity of $G$.} \label{fig:reduction}	
	\end{figure*} 
	
	Our  analysis begins with the complexity of answering the  question Q2: \emph{is there a schedule that satisfies at least $m$ users?}
	
	To prove our complexity theorem we will make use of the notion of independent sets on graphs. Consider an undirected graph $\Gc=(\Vc,\Ec)$ with vertices $\Vc$ and edges $\Ec$, such as the one in the example of Fig.~\ref{fig:reduction}-(b). 
	\begin{definition}
		A subset of nodes $I\subseteq \Vc$ is called an independent set on graph $\Gc=(\Vc,\Ec)$ if  any two nodes $v,u\in I$ are not neighbors on $\Gc$, 
		i.e. $\{v,u\}\notin \Ec$.
	\end{definition}
	For example, one may verify that the set of nodes $\{2,4,5\}$ forms an independent set in the graph of Fig.~\ref{fig:reduction}-(b).
	From the literature we know  that finding the maximum independent set in a graph is NP-hard, and its decision version ``\emph{is there an independent set of size at least $m$?}'' is
	NP-complete \cite{Austrin11}.
	
	Further, the connectivity of a graph can be represented in an alternative way,  shown in Fig.~\ref{fig:reduction}-(c). Specifically, we may consider a bipartite graph $\Hc(\Gc)=(\Vc\cup\Ec,\Lc)$ which is built from $\Gc$ as follows. The left node partition is set to $\Vc$ and the right node partition is set to $\Ec$, hence the nodes of $\Hc$ is the set $\Vc\cup\Ec$. Then $(v,e)\in\Lc$ if and only if $v\in e$. 
	An independent set on $\Gc$ is a selection of  left nodes of $\Hc$ such that the induced graph (formed by keeping the selected left nodes, the right nodes, and the surviving links) has right degree at most 1.

	\begin{theorem}\label{th:basicHardnessThm}
		It is NP-complete to determine whether there exists a feasible URLLC  schedule for $|\Mc|\geq m$ users (i.e. to answer Q2) in the binary SNR model. 
	\end{theorem}
	\begin{IEEEproof}
		We will establish a reduction from the decision problem \emph{is there an independent set of size at least $m$?}, which is  NP-complete.
		

		For an instance of Q2  in the binary model we may  construct the following URLLC bipartite graph connecting users and RBs, where we draw a link from  user $k$ to RB $r$ if  $\delta_r^k=1$ in the binary model (i.e. the RB is active for this user),~cf.~Fig.~\ref{fig:reduction}-(a). 
		In order to perform the reduction, we first assume that there exists a \emph{URLLC oracle} algorithm that given input $(\Kc,\Rc,\boldsymbol\delta,d_k, m)$\footnote{We remind that, $\mathcal{K}$ stands for the set of all URLLC users, $\mathcal{R}$ stands for the set of  all resource blocks, $\boldsymbol\delta$ is a matrix with $\delta_r^k=1$ denoting that block $r$ is ``active'' for user $k$,$d_k$ is the number of required successful transmissions for the satisfaction of the URLLC constraint and $m$ is the number of users we want to satisfy.} 
		it provides the answer to Q2
		in polynomial time. The reduction is to show that we may use this algorithm  to solve every instance of the independent set decision problem. 
		
		Consider any graph $\Gc=(\Vc,\Ec)$, and the decision problem  \emph{``is there an independent set of size at least $m$?''} on the corresponding bipartite connectivity graph $\Hc(\Gc)=(\Vc\cup\Ec,\Lc)$. 
		Run the URLLC oracle with $\Kc=\Vc$, $\Rc=\Ec$,  $\delta_e^v=1$ if and only if $v\in e$ ($e$ is incident to $v\in\Vc$ in $\Gc$), and $d_v=M_{v}$, where $M_v$ is the degree of node $v$ in $\Gc$. Report the answer of the oracle as the answer to our decision problem.
		
		First, the above procedure runs in polynomial steps by the hypothesis of the  oracle. To prove the correctness, we work as follows. Suppose that the URLLC oracle encounters a subset of users $\Uc\subseteq\Vc$ that are found SLA feasible, then there exists an activation of links (feasible schedule) such that  \emph{(i)} for every $v\in \Uc$, all $M_v$ links are activated (i.e. all $d_v$ transmissions are scheduled), and \emph{(ii)} the number of activated links incident to any right node is $\leq 1$ (since the schedule is feasible); it follows that   $\Uc$ is an independent set on $\Gc$. We conclude that \emph{the URLLC oracle finds a set $\Uc$ to be SLA feasible  if and only if $\Uc$  forms and independent set on $\Gc$}.
		Hence, if the oracle returns ``yes'', then we know there exists an independent set of size $m$ in $\Gc$. Conversely, if it returns ``no'', we know that there is no  independent set of size $m$ or larger in $\Gc$. This completes the reduction.	
		The problem is NP-complete because, as we will show next,  given a candidate solution $\Mc$ we may verify its feasibility in polynomial time, hence our problem is in NP.
		%
	\end{IEEEproof}
	\begin{corollary}\label{cor:2}
		The UUM problem in \eqref{eq:maxutil} is NP-hard. 
	\end{corollary} 
	Corollary \ref{cor:2} can be proven by a reduction from Q2. Suppose UUM can be solved in polynomial time for any instance, then select an instance with $w^k=1,~\forall k$, and run the UUM oracle. The obtained maximum utility can be used to directly determine  the answer to Q2. That is, UUM is no easier than Q2, which, in turn, is no easier than the independent set decision problem.
	
	\begin{corollary}
		In the continuous SNR model obtained by replacing \eqref{eq:cstr1} with $p_e^k(\xv) +\theta -1 \leq 1-z^k$ and omitting \eqref{eq:cstr255}, Q2 and  UUM are NP-hard. 
	\end{corollary} 
	
	In the continuous SNR model, consider instances that have only two possible values for SNRs, 0 and the threshold used for the binary SNR model. The arising instances coincide with  respective instances in the binary SNR model, hence all instances of the binary model also appear in the continuous, hence also the hard ones.
	

	Finally, we can use the independent set construction above to bound the approximation ratio of the UUM problem in the binary SNR model:
	\begin{proposition}\label{prop:APXhard}
		Denote $d = \max_k[d_k]$, and consider the case $w^k=1,~\forall k$. If $P\!\neq\! NP$ and the ``Unique Games Conjecture''\footnote{For more details about this conjecture cf.~\cite{Khot2005}} holds, the UUM problem under the binary SNR model admits no polynomial time algorithm with approximation ratio better than $\frac{(\log_2d)^2}{d}$. 
	\end{proposition}
	\begin{IEEEproof}
		Indeed, notice that $d = \max_k[d_k]$ is the maximum left degree  of the bipartite graph, which is $d_{max}(\Gc)$ when transforming a general graph $\Gc$ to it, as described in the proof of Theorem \ref{th:basicHardnessThm}. Therefore, if a better approximation was possible in polynomial time, that oracle could be used to obtain in polynomial time an independent set that approximates the optimal  better than $\frac{(\log_2d)^2}{d}$, which, if the Unique Games Conjecture holds, is not possible unless $P\!=\!NP$ \cite{Austrin11}. 
	\end{IEEEproof}

	\section{Optimal URLLC Scheduling in  Binary SNR}\label{sec:effsch}
	
	Having established that answering Q2 and solving the UUM problem are both very complex, in this Section we shift our attention to the scheduling problems. Surprisingly, we will prove in the binary SNR model that given a designated set of users $\Mc$,  we can answer if the set is schedulable (Q1) and find a feasible schedule (when the answer is ``yes'') in polynomial time. In turn, this result is very important as \emph{(i)} it allows us to achieve maximum  URLLC scheduling performance with low-complexity algorithms, and \emph{(ii)}  will lead us to obtain an efficient admission control algorithm.
	
	Our analysis is based on the concept of Total Unimodularity of a matrix. Formally we have: 
	
	\begin{definition}[Total Unimodularity] 
		A (square) matrix  $\Bm$ is called unimodular if $\det(\Bm)\in\{-1, 0, 1\}$. A matrix $\Am$ is called totally unimodular if every square submatrix of $\Am$ is unimodular.
	\end{definition}
	
	\noindent This concept is of great importance in Linear Programming.
	Let the polyhedron $\Pc(\Am, \bv)=\{\xv : \Am\xv\leq \bv, \xv\geq 0\}$ be the feasible set of a Linear Program (LP). If a matrix $\Am$ is totally unimodular and $\bv$ a vector of integers, then
	the vertices of $\Pc(\Am, \bv)$ have all integral elements  \cite[Theorem 13.2]{Papadimitriou_book}. This implies that any LP of the form $\max_{\xv\in\Pc(\Am, \bv)}\cv^T\xv$
has at least one integral solution, and if its solution is unique, then it is necessarily integral. 
	
	Our strategy will be   (i) to show that an appropriate relaxation of the UUM is an LP with  a totally unimodular constraint matrix and (ii) to construct a mock objective function such that the corresponding LP has unique solution.  
	
	Recall that we are given a set of users $\Mc$ and we want to decide if there exists an $\Mc$-SLA feasible schedule, that is if set $\Xc(\Mc)$ is nonempty. 
	As a first step, starting from the feasibility space of \eqref{eq:maxutil}, we can fix $z^k\!=\!1, \text{if }k\in\Mc$ and 0 otherwise, to arrive at an expression for $\Xc(\Mc)$:
	\[
	\Xc(\Mc)=\left\{\xv \in\{0,1\}^{R\times |\Mc|} ~\Bigg|~ \begin{array}{ll}
	\sum_{r}x_r^k \geq d_k, &  k\in\Mc\\
	\sum_{k}x_{r}^k \leq 1, & r\in\Rc\\
	x_r^k \leq \delta_r^k,& \forall (r, k) \label{eq:cstr25}\\
	\end{array} \right\}
	\]
	The idea is that we will relax the scheduling variables to $\xv\in[0,1]^{R\times |\Mc|}$ and obtain an LP. In order to force the LP to have a unique solution, we introduce random costs $c_r^k$, which are drawn uniformly from $[0,1]$.  We then have: 
	
	\vspace{0.1in}
	\noindent \emph{\underline{Relaxed LP} }
	\begin{equation}
	\min_{\xv\in\Xc(\Mc)} \sum_{r=1}^R\sum_{k=1}^{|\Mc|}c_{r}^{k}x_r^k \label{eq:relaxedLP} 
	\end{equation}
	
	\begin{lemma}\label{lem:unimodularity_relaxed_lp}
		The constraint matrix of the relaxed LP  \eqref{eq:relaxedLP} is totally unimodular. 
	\end{lemma}
	\begin{IEEEproof}
		First, let us stack all variables in vector
		\[
		\xiv = [x_1^1, x_1^2, \dots, x_1^M, x_2^1,\dots, x_2^M, \dots, x_R^1, \dots x_R^M]^T  
		\]
		The constraints $\xv \in \Xc(\Mc)$ of the relaxed LP then have the form $\Am\xiv \leq \bv, \xiv \geq \zerov$, with constraint matrix\footnote{We use the notation $\pmb{I}_N$ for the identity matrix of size $N$.} 
		\[
		\Am = \left[\Am_1^T \quad \Am_2^T \quad \pmb{I}_{RM} \right]^T
		\],
		where:
		\begin{itemize}
			\item $\Am_1$ is the $M\times RM$ matrix corresponding to the first set of constraints, i.e. $\Am_1=-[\pmb{I}_M| \pmb{I}_M|...|\pmb{I}_M]$, where the identity matrix of size $M$ appears $R$ times.  
			\item $\Am_2$  is the $R\times RM$ matrix corresponding to the second set of constraints, i.e. its $r$-th row has elements in columns $\{(r-1)M+1, (r-1)M+2,...,(r-1)M+M\}$ equal to one and the rest zero. 
		\end{itemize}
		
		We can thus observe that all of the following are true for the matrix $\Am_3= [-\Am_1^T  \quad \Am_2^T]^T$:
		
		\begin{enumerate}
			\item Every entry is either $0$ or $1$. 
			\item At each column, there two nonzero elements, both taking value $1$.
			\item Sets $\Sc_1, \Sc_2$ that have as elements the rows of  $\Am_1$ $\Am_2$, respectively are disjoint. 
			\item For every column one of the rows with a nonzero element belongs to $\Sc_1$ and the other belongs to $\Sc_2$. 
		\end{enumerate}
		It then follows \cite[Th. 13.3]{Papadimitriou_book} that the matrix $\Am_3$ is totally unimodular. Since multiplying rows of a totally unimodular matrix by $-1$ results in a totally unimodular matrix, the matrix $\Am_4 = \begin{bmatrix}
		\Am_1^T & \Am_2^T
		\end{bmatrix}^T $
		is totally unimodular, therefore the constraint matrix $
		\Am = \begin{bmatrix}
		\Am_4^T &	\pmb{I}_{RM}
		\end{bmatrix}^T 
		$ is totally unimodular as well, 
		completing the proof. 
	\end{IEEEproof}
	
	We now present Algorithm 1, which uses the relaxed LP to answer Q1. It has to be noted that solving the relaxed LP here means to run a procedure which returns an optimal solution if the LP is feasible and an indication that is infeasible otherwise, which can be done, for example, using the Ellipsoid algorithm, see for example~\cite[Chapter 8]{Papadimitriou_book}. 
	
	\begin{algorithm}
		\caption{Check feasibility in the binary SNR model}
		\label{alg:feasibilityChecker}
		\begin{algorithmic}
			\STATE Construct the relaxed problem.\\
			\STATE Select $c_{r,k}\in[0,1]$ uniformly at random.\\ 
			\STATE Solve the relaxed LP \eqref{eq:relaxedLP}.\\
			\IF{The LP is feasible} 
			\STATE \textbf{Return:} (yes) $\Mc$ is feasible\\
			\STATE \textbf{Return:} The solution $\xv^*$ of the LP as a schedule.
			\ELSE
			\STATE \textbf{Return:} (no) $\Mc$ is not feasible
			\ENDIF
		\end{algorithmic}
	\end{algorithm}
	
	\begin{theorem}\label{th:LPRelaxationThm}
		Algorithm 1 always returns a correct answer to Q1, and with probability $1$ a feasible schedule  if the answer is ``yes".  
	\end{theorem}
	\begin{IEEEproof}
		If the relaxed LP is infeasible, we may immediately conclude that  there is no feasible schedule for $\Mc$ users. On the other hand, if the LP is feasible  then since from Lemma \ref{lem:unimodularity_relaxed_lp} its corresponding constraint matrix is totally unimodular and the right hand sides of the constraints are integers, at least one solution should be integral (see \cite[Theorem 13.2]{Papadimitriou_book}), therefore a feasible schedule for $\Mc$ users.  We can then conclude that Algorithm 1 returns a correct answer to Q1. 
		
		Assume now that the relaxed LP is feasible. Since the cost vector $\cv$ is chosen uniformly at random, the probability that the hyperplane $\cv^T\xv=0$ is parallel with any of the facets of the LP polyhedron is zero, therefore the relaxed LP has a unique solution with probability one. Due to total unimodularity of the constraint matrix, the unique solution obtained in this way is necessarily integral, therefore a feasible schedule. 	
	\end{IEEEproof}

	An immediate corollary is that Q1 can be answered in polynomial time: 
	\begin{corollary}
		Question Q1 can be answered in polynomial time (in $R, |\Mc|$ and $N$) for the binary SNR model. 
	\end{corollary}
	\begin{IEEEproof}
		The relaxed problem has $R|\Mc|$ variables and $R+|\Mc|+2R|\Mc|$ constraints, which are both polynomial in $R$ and $|\Mc|$. Hence Q1 can be answered by checking the feasibility and finding the solution of an LP with size polynomial to $R$ and $\Mc|$, which can be done in polynomial time, e.g. with the Ellipsoid algorithm \cite[Chapter 8]{Papadimitriou_book}.  
	\end{IEEEproof}
	\section{Approximate Admission Control in Binary SNR}\label{sec:approx_bin}	

	In this section, we prove that the \textup{GREEDY} algorithm guarantees  $1/(d+1)$ of the maximum in the joint admission control and scheduling problem, where $d$ refers to the maximum number of active RBs required among users.  For example, in case $d_k=d=3$ (as in the introduction), then this shows that \textup{GREEDY} achieves at least $25\%$ of the optimal.
	Following proposition \ref{prop:APXhard}, we may conclude that \textup{GREEDY} achieves the  optimal approximation  up to poly-logarithmic terms. Hence, we can not hope for a much better approximation guarantee. 

	The GREEDY algorithm works as follows: \emph{(i)} The users are ordered in decreasing utilities $w^{(1)}\geq \dots\geq w^{(K)}$ with ties broken randomly, \emph{(ii)} starting from highest utilities we allocate to $k$ user  $d_k$ RBs at random, \emph{(iii)} if there are not enough RBs, then the user is rejected altogether. 
	
	\begin{proposition}\label{prop:approxRatioBinarySNR}
		Let $d = \max_{k\in\Kc}d_k$, $w^k=1,\forall k$. 
		Then \textup{GREEDY} guarantees an approximation ratio of at least $\frac{1}{d+1}$ for the binary SNR model. 
	\end{proposition}
	\begin{IEEEproof}
		We assume, without loss of generality, that $\sum_{r}\delta_{r}^{k} \geq d_k,~\forall k\in\Kc$, i.e. each user has enough active resource blocks (if not we may eliminate those users and redefine $\Kc$). Let $\hat{\xv}$ denote the schedule returned by GREEDY, $\hat{\zv}$ the corresponding admission and $\zv_*$ the optimal admission. 
		We note that  if user $k$ is admitted in GREEDY it gets allocated exactly the minimum required number of resource blocks and if not it is allocated no resource blocks at all.  
		
		
		We may partition the sets of users and resource blocks into two (disjoint) subsets: $\Kc_1, \Rc_1$ are  the admitted users  and assigned resource blocks by GREEDY respectively,  while $\Kc_0=\Kc\setminus \Kc_1, \Rc_0=\Rc\setminus\Rc_1$. The following are true: \emph{(i)} $|\Rc_1| \leq d |\Kc_1|$, since each user needs at most $d$ resource blocks and \emph{(ii)} no user in $\Kc_0$ can be scheduled successfully with only RBs from $\Rc_0$ (by the premise that they are not scheduled by GREEDY). 
		
Since no user in $\Kc_0$ can be scheduled with RBs exclusively from $\Rc_0$,
		we observe that an upper bound on the admissible users (hence on $\sum_{k}z_*^k$) is $|\Rc_1|+|\Kc_1|$, which is attained if all users in $\Kc_1$ can be scheduled exclusively with  $\Rc_0$ RBs and $|\Rc_1|$ users from $\Kc_0$ can be scheduled with 1 RB from $\Rc_1$ and the rest from $\Rc_0$. Therefore, we have 
		\[
		\sum_{k}z_*^k\leq |\Rc_1|+|\Kc_1| \leq d|\Kc_1| + |\Kc_1| = (d+1)\sum_{k}\hat{z}^k
		,\]
		finishing the proof.   
	\end{IEEEproof}

	We further mention that in case $d=1$, and hence $d_k=1,~\forall k$, then the
	URLLC joint admission control and scheduling simplifies to a maximum weighted matching problem, which can be solved optimally in polynomial time, see e.g. \cite[Chapter 10]{Papadimitriou_book}. Furthermore, if there exists a $\Kc$-SLA  feasible schedule, we can find it in polynomial time by running Algorithm \ref{alg:feasibilityChecker} before GREEDY, thus the performance is optimal in this case as well.  
	
	\begin{figure}[t!]
		\centering
		\includegraphics[width=0.48\linewidth]{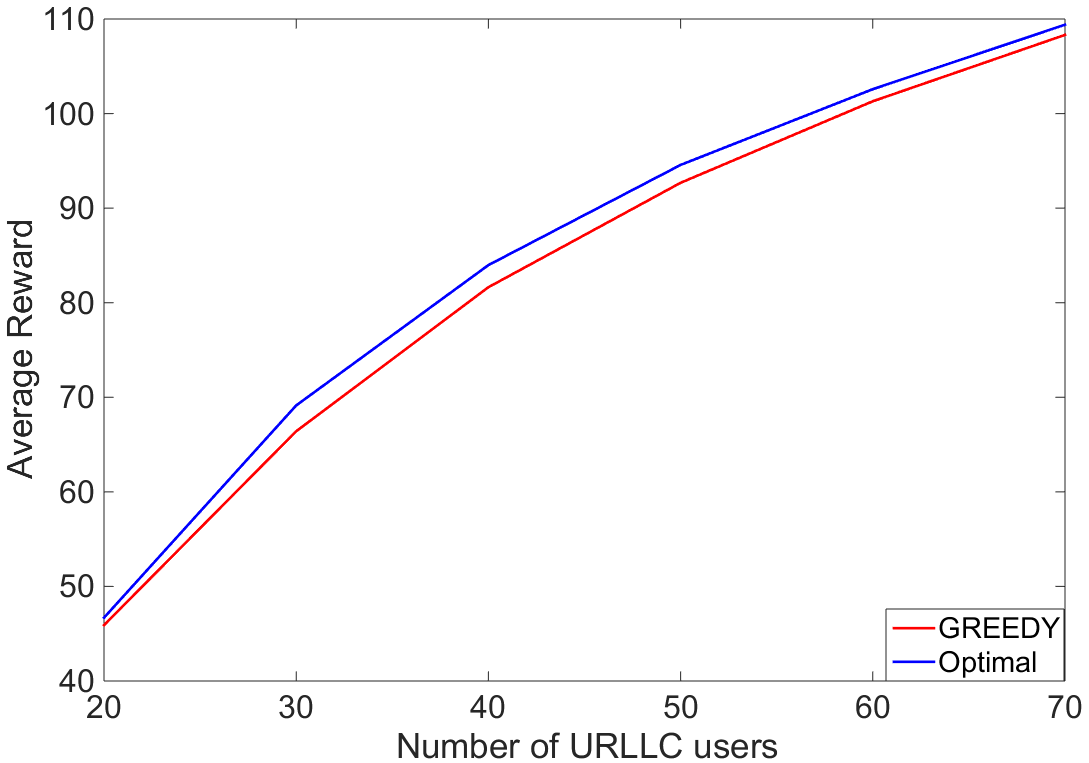}\hspace{0.02\linewidth}
		\includegraphics[width=0.48\linewidth]{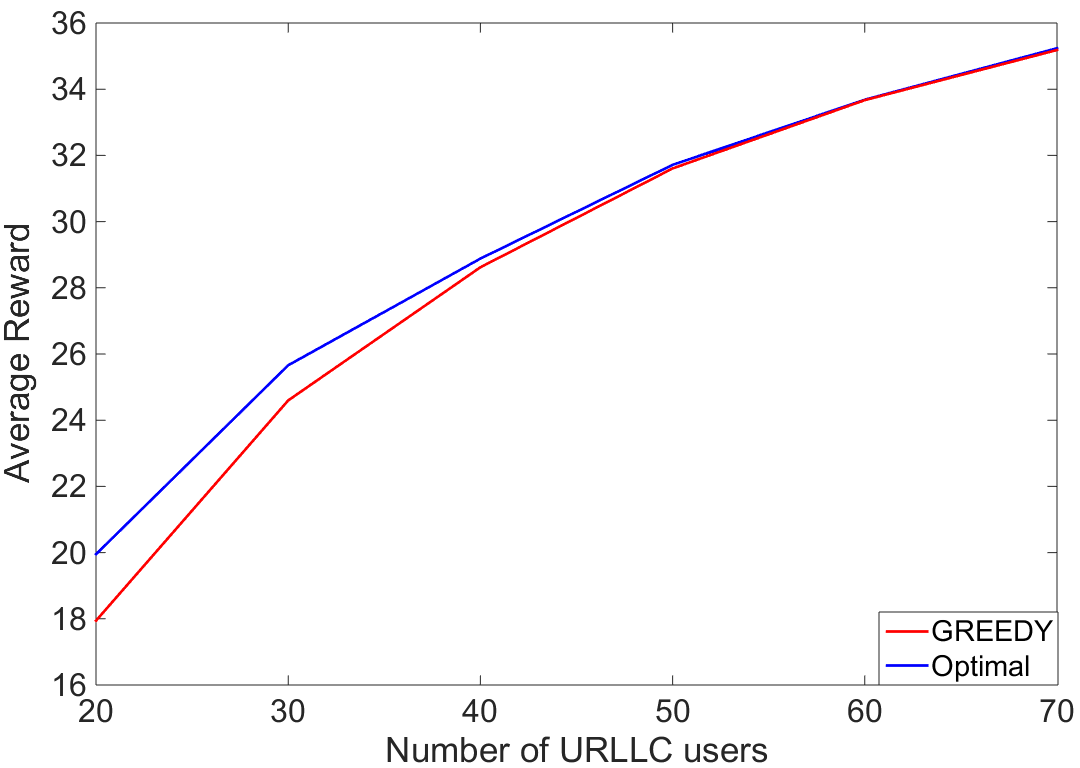}%
		\caption{Performance comparison (each point is the average over $10000$ trials) between GREEDY and optimal in the binary SNR model. (left) Each user has a random utility in $[0, 5]$. (right) All users have the same utility.}%
		\vspace{-0.2in}
		\label{fig:binarySNRperformance}%
	\end{figure} 
	Figure \ref{fig:binarySNRperformance} shows a performance comparison between GREEDY and the optimal solution of the UUM problem, which was obtained via an ILP solver. We perform the comparison for different simulation settings.  
	Users are placed at random in an area such that their mean SNRs computed using path loss models are between $0$ and $20$ $dB$. The system has $50$ resource blocks, each experiencing $i.i.d.$ Rayleigh fading. The thresholds are chosen such that $d_k=1$ if the average SNR of user $k$ is over $12.5dB$, $d_k=3$ if it is lower than $4dB$ and $d_k=2$ in between. In the left case, each user has a random utility in $[0,5]$, while in the right all users have the same utility.
	
	We can observe that in all cases the gap between the performance of the two algorithms is very small (in fact GREEDY performs much closer to optimal than the predicted guarantee), therefore validating our intuition that GREEDY is an effective solution for the admission control problem the binary SNR model. In addition, GREEDY is a  fast algorithm compared to the one that solves the ILP optimally; on average, the runtime of GREEDY was about $25$ times faster than solving the ILP exactly. 
	
	Next, we will use the GREEDY algorithm together with the result of Section \ref{sec:complexity} that the feasibility problem can be solved in polynomial time for the binary SNR model to provide an admission controller for the continuous SNR model.

	\section{Admission Control in Continuous SNR}\label{sec:approx_con}

	We now shift our attention  to the continuous SNR model, where joint coding is performed over the resource blocks that belong to the same user. As we discussed in Section \ref{sec:complexity}, the binary SNR model is a special instance of this general case, therefore all hardness results hold here as well.  In addition, even the fractional relaxation of the problem is difficult to address, since it is in fact a non-convex problem.  To see this, recall that the corresponding probability of incorrect decoding is given by \eqref{eq:errorProb}. In order to satisfy the SLA constraint we should have, using also \eqref{eq:cstr1} that 
	\begin{align}\nonumber 
	\sum_{r}(n\log_2(1+&\gamma_r^k) + 0.5\log_2(n))x_r^k \\ \nonumber 
	&- Q^{-1}(1-\theta)\sqrt{n\sum_{r}V(\gamma_r^k)x_r^k} -L_k \geq 0, \forall k
	.\end{align} 
	Since the left hand side is a  convex function of $\xv$ and is required to be greater than $0$, the points satisfying these constraints form a non-convex set. 
	
	A baseline greedy approach to solve this problem is to order the users in decreasing utilities, and then place one by one at random RBs if their constraint is satisfied. Below we propose a binary SNR-inspired heuristic, called \emph{Iterative Thresholding Algorithm} (ITA) and show via experimentation that outperforms the mentioned baseline. 
	
	To design ITA, we have used the following insights gained from the analysis of the binary SNR model. Namely, we leverage the facts that \emph{(i)} only a few resource blocks are needed for each user as illustrated by Fig. \ref{fig:resourceBlocksNeeded},  \emph{(ii)} if a feasible schedule exists for the binary SNR model, we can obtain it in polynomial time as we showed in Section \ref{sec:approx_bin} and \emph{(iii)} GREEDY gets close to the optimal as we showed in Section \ref{sec:approx_bin}. Our idea is to start with a high SNR threshold, use the corresponding binary SNR model for this threshold to assign RBs via Alg. \ref{alg:feasibilityChecker} if all users are schedulable (or via GREEDY otherwise), fix the satisfied users and their assigned RBs, and  then progressively lower the SNR threshold repeating the same procedure. The proposed algorithm is detailed as Algorithm \ref{alg:iterativeThreshold}.
	
	\begin{algorithm}
		\caption{Iterative Thresholding Algorithm (ITA)}
		\label{alg:iterativeThreshold}
		\begin{algorithmic}
			\STATE Initialization: $\Mc\leftarrow \Kc, \Sc \leftarrow\emptyset, \xv \leftarrow \zerov$. 
			\FOR {$d=1,2,...,D_{max}$} 
			\STATE Find the minimum SNR $s(d)$ such that $d$ resource blocks of SNR $s(d)$ are sufficient.\\
			\STATE For each user $k\in\Mc$ and resource block $r\in\Rc$, put $\delta_{(r,k)} = \ind{\gamma_r^k \geq s(d)}$
			\STATE Run Algorithm 1 for users in $\Mc$ and blocks in $\Rc$. 
			\IF {the problem is feasible}
			\STATE Return $\xv(d)$ as the feasible schedule. 
			\ELSE
			\IF{$d=1$}
			\STATE Run a maximum weighted matching algorithm in the resulting connectivity graph, return $\xv(d)$ as the resulting schedule.
			\ELSE
			\STATE Run GREEDY for users in $\Mc$ and blocks in $\Rc$, return $\xv(d)$ as the resulting schedule.
			\ENDIF 
			\ENDIF  
			\STATE Update the schedule: $\xv \leftarrow \xv + \xv(d)$.
			\STATE Update the set of scheduled users: $\Sc\leftarrow \Sc\cup\left\{k\in\Mc:\sum_r\xv_{r}^k(d)\geq d \right\}$
			\STATE Update the set of remaining users: $\Mc \leftarrow \Kc \setminus \Sc$. 
			\STATE Update the set of available resources: $\Rc \leftarrow \Rc \setminus \{r\in\Rc : \sum_{k}x_r^k >0\}$ 
			\IF {$\Rc=\emptyset$}
			\STATE Break from the loop 
			\ENDIF
			\ENDFOR
			\STATE Return the  schedule $\xv$. 
		\end{algorithmic}
	\end{algorithm}
	
\begin{figure*}[t]
		\centering
		\subfigure[$10$ Resource Blocks]
		{\includegraphics[width=0.32\linewidth]{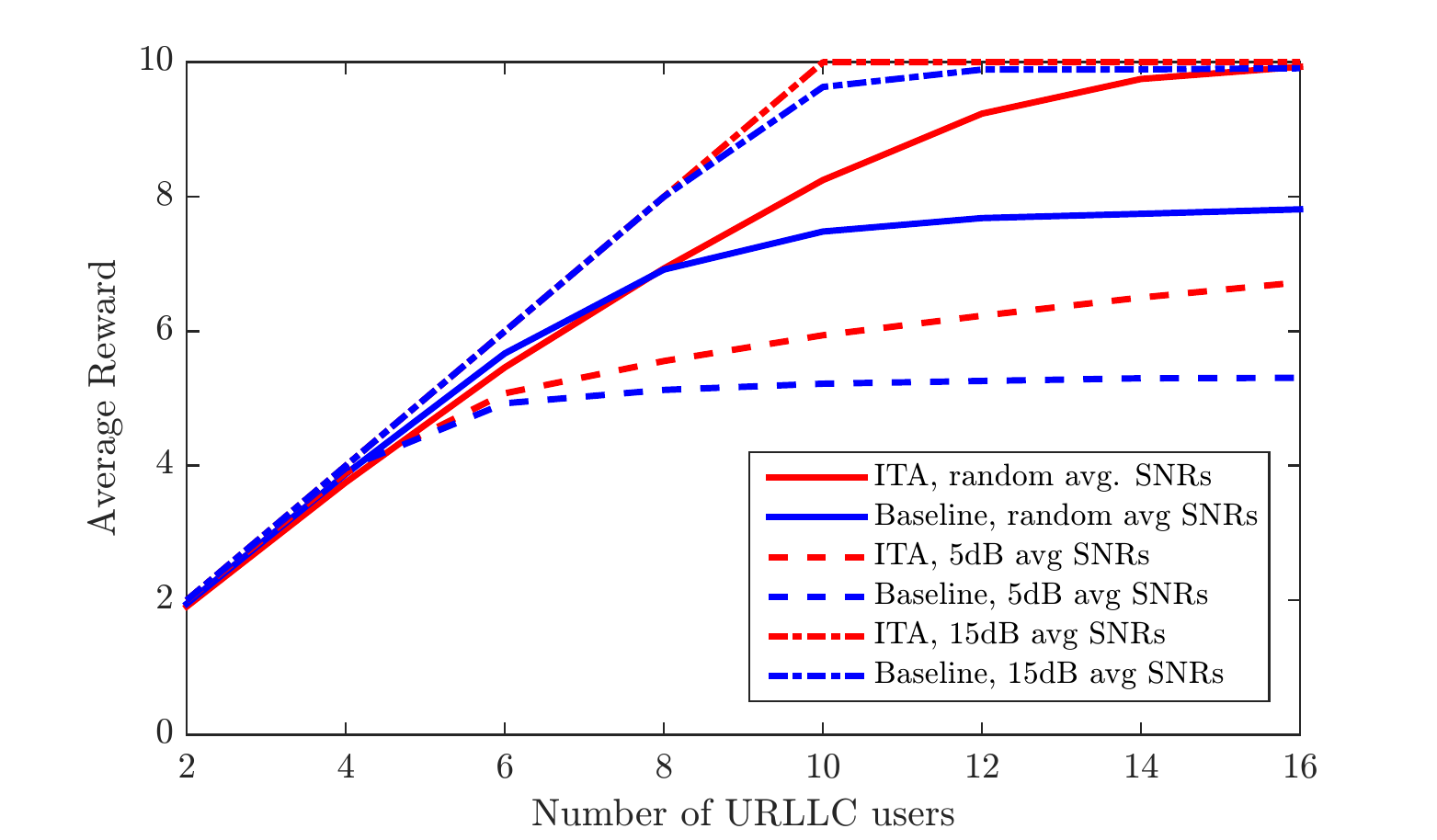}
		}
		\subfigure[$30$ Resource Blocks]
		{\includegraphics[width=0.32\linewidth]{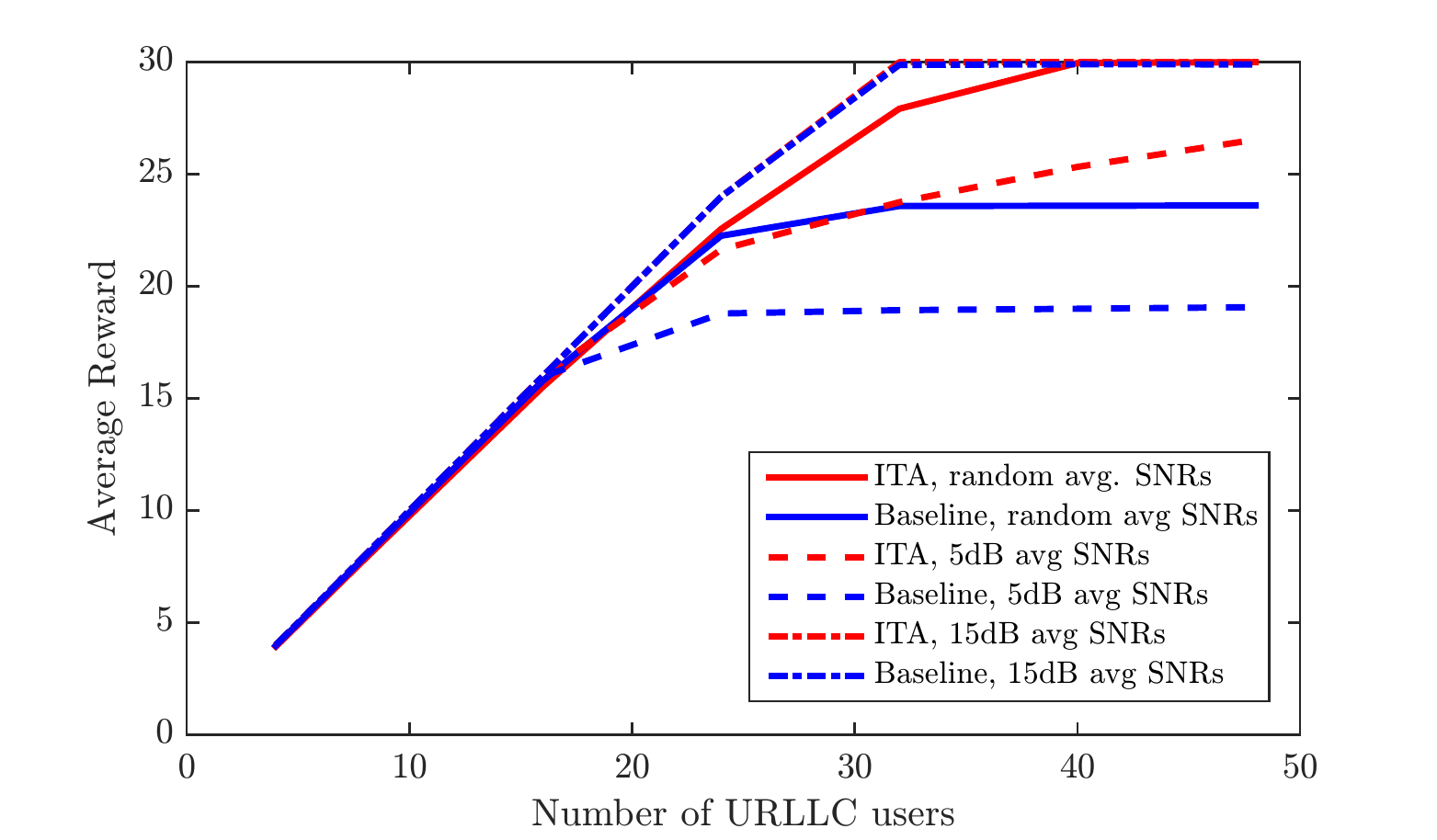}
		}
		\subfigure[$60$ Resource Blocks]
		{\includegraphics[width=0.32\linewidth]{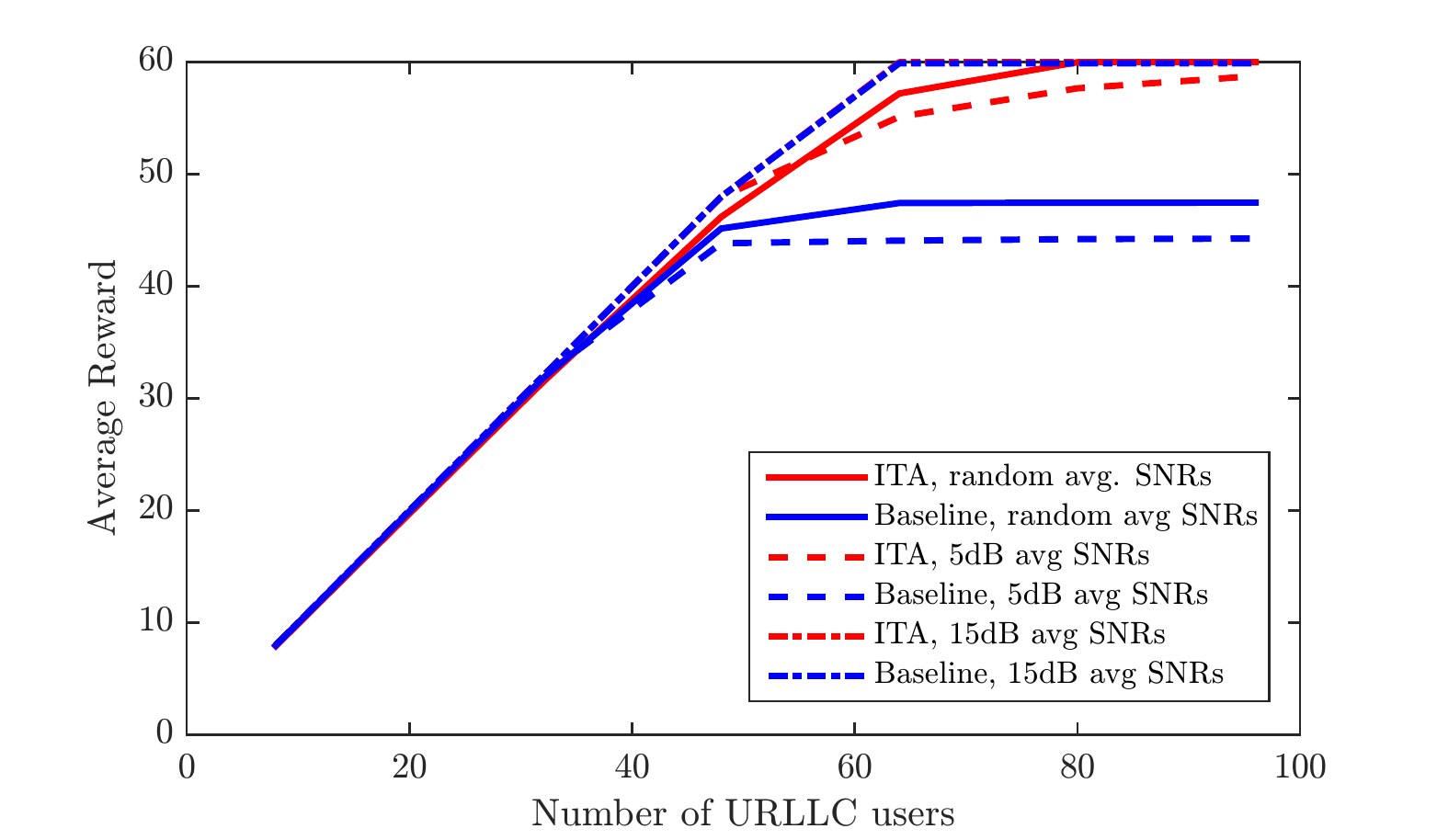}
		}
		\caption{Average number of admitted URLLC users (here admitting each user has unit reward) with (a) low (b) medium and (c) high number of resources allocated to URLLC traffic. Each user has a packet size of $32${ Bytes } and requests $99.999\%$ reliability. The results shown are  averages over $5000$ trials with generated Rayleigh fading.} 
		\label{fig:sameRewards}
	\end{figure*}

	We compare Alg. \ref{alg:iterativeThreshold} to the baseline explained above. Results regarding the problem of maximizing the number of admitted users (i.e. $w^k=1$ for every $k$) are shown in Fig. \ref{fig:sameRewards}. We examine three cases regarding user placement: (i) users are placed at random in the cell with mean SNRs (due to large scale fading) between $0dB$ and $20dB$ and the cases where users have the same mean SNR with a (ii) relatively low ($5dB$) and (iii) relatively high ($15dB$) value. We can observe that where the SNR is high (i.e. users are placed close to the Base Station) the performance of the two algorithms is almost the same. Moreover, the two algorithms admit similar number of users in all cases where the number of RBs is much lower than the total number of URLLC users. These results were to be expected since in the latter case most resource blocks are good for every user and in the former there are enough resources to find good RBs for each user. More interestingly though, when the number of RBs is comparable and/or lower to the number of URLLC users and the users are placed at random or have the same and relatively low mean SNRs--the regime where admission control really becomes an important aspect of the problem--\emph{ITA outperforms the baseline algorithm by a margin that increases with the number of users and can reach up to around $30\%$}. The reason behind this gain is that ITA schedules efficiently groups of users for each threshold, while the baseline  algorithm may use a resource block that is more useful to some other user.  

	\begin{figure*}[t]
		\centering
		\subfigure[$10$ Resource Blocks]{\includegraphics[width=0.32\linewidth]{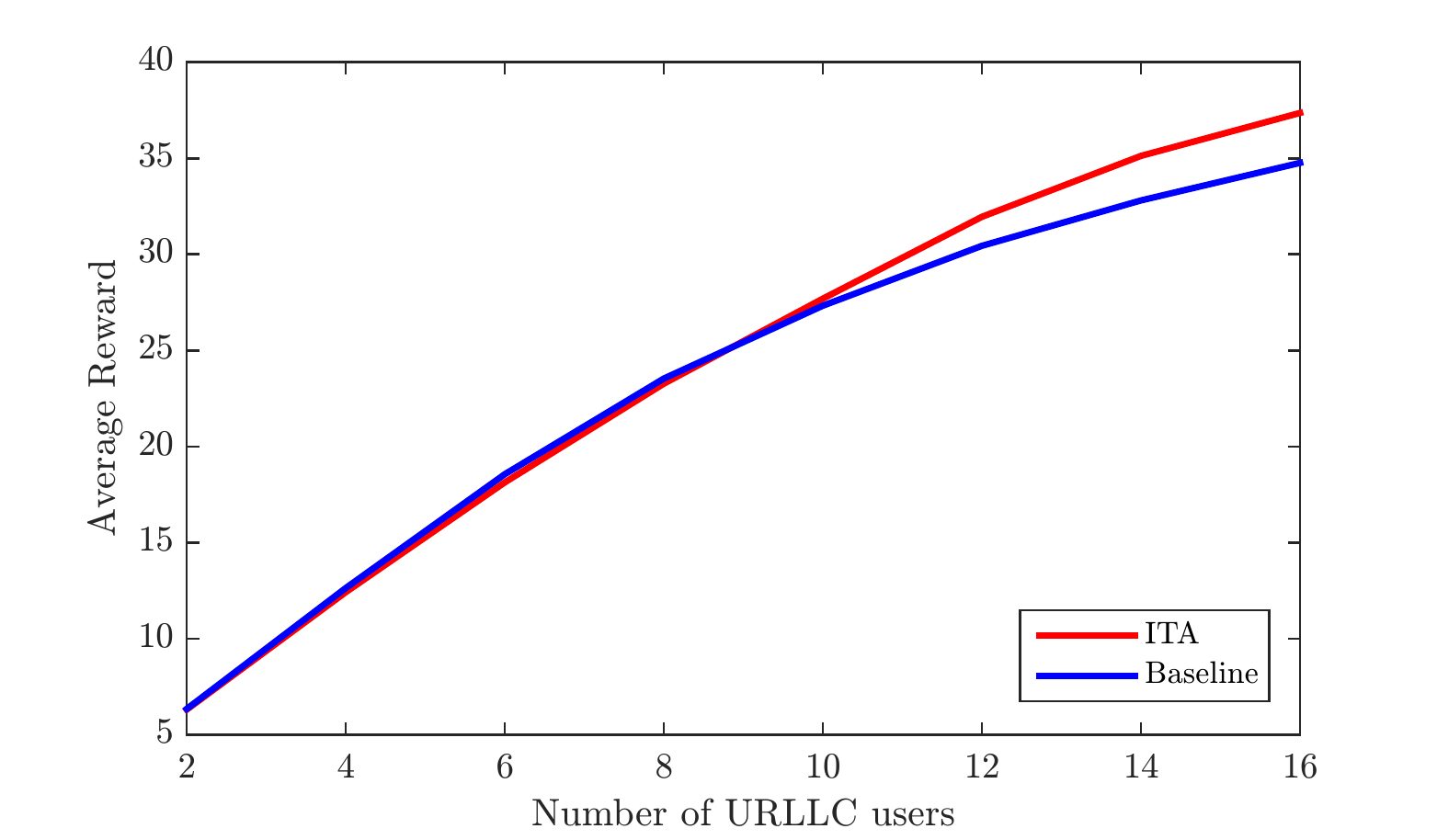}
		}
		\subfigure[$30$ Resource Blocks]{\includegraphics[width=0.32\linewidth]{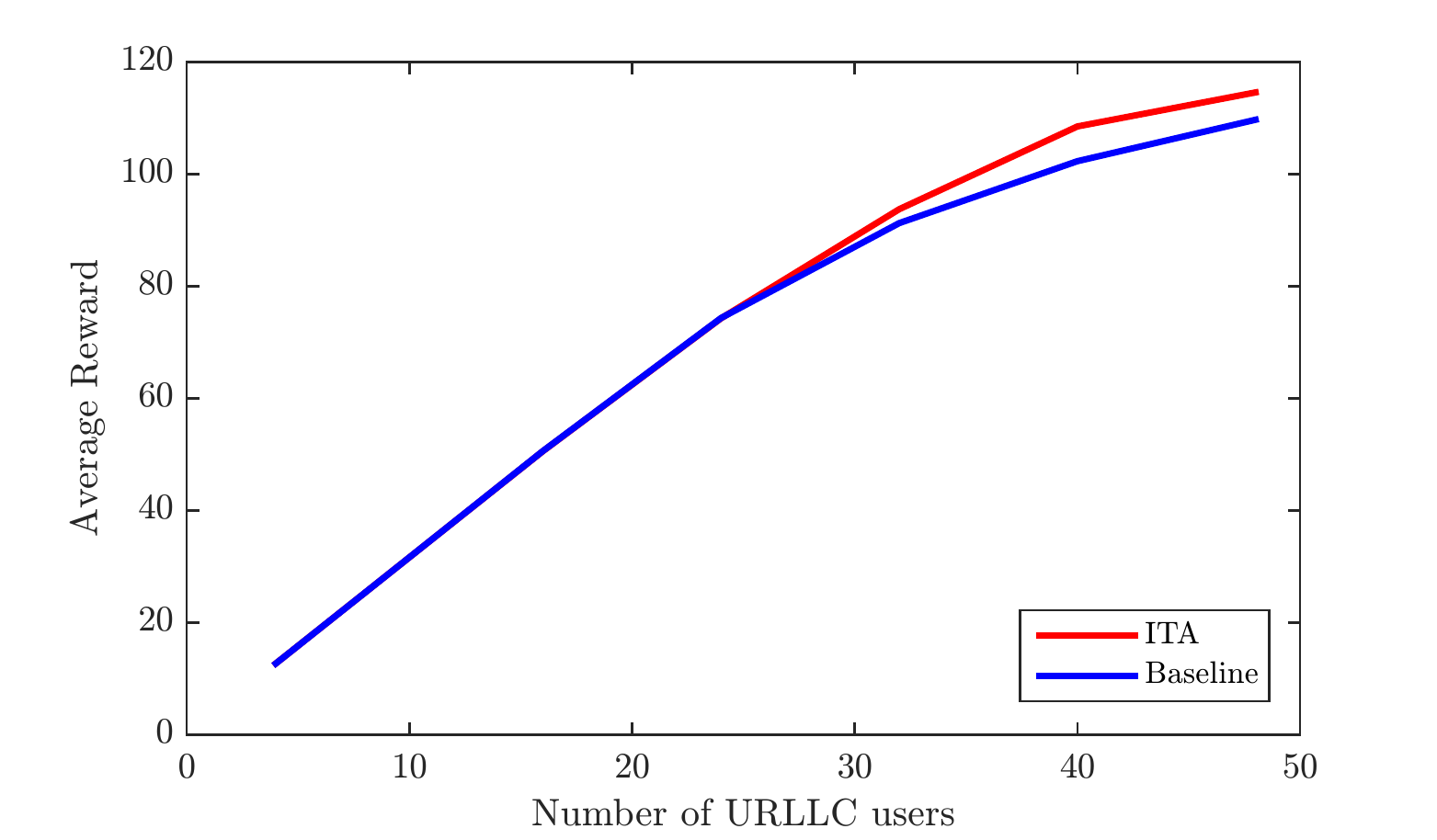}
		}
		\subfigure[$60$ Resource Blocks]{\includegraphics[width=0.32\linewidth]{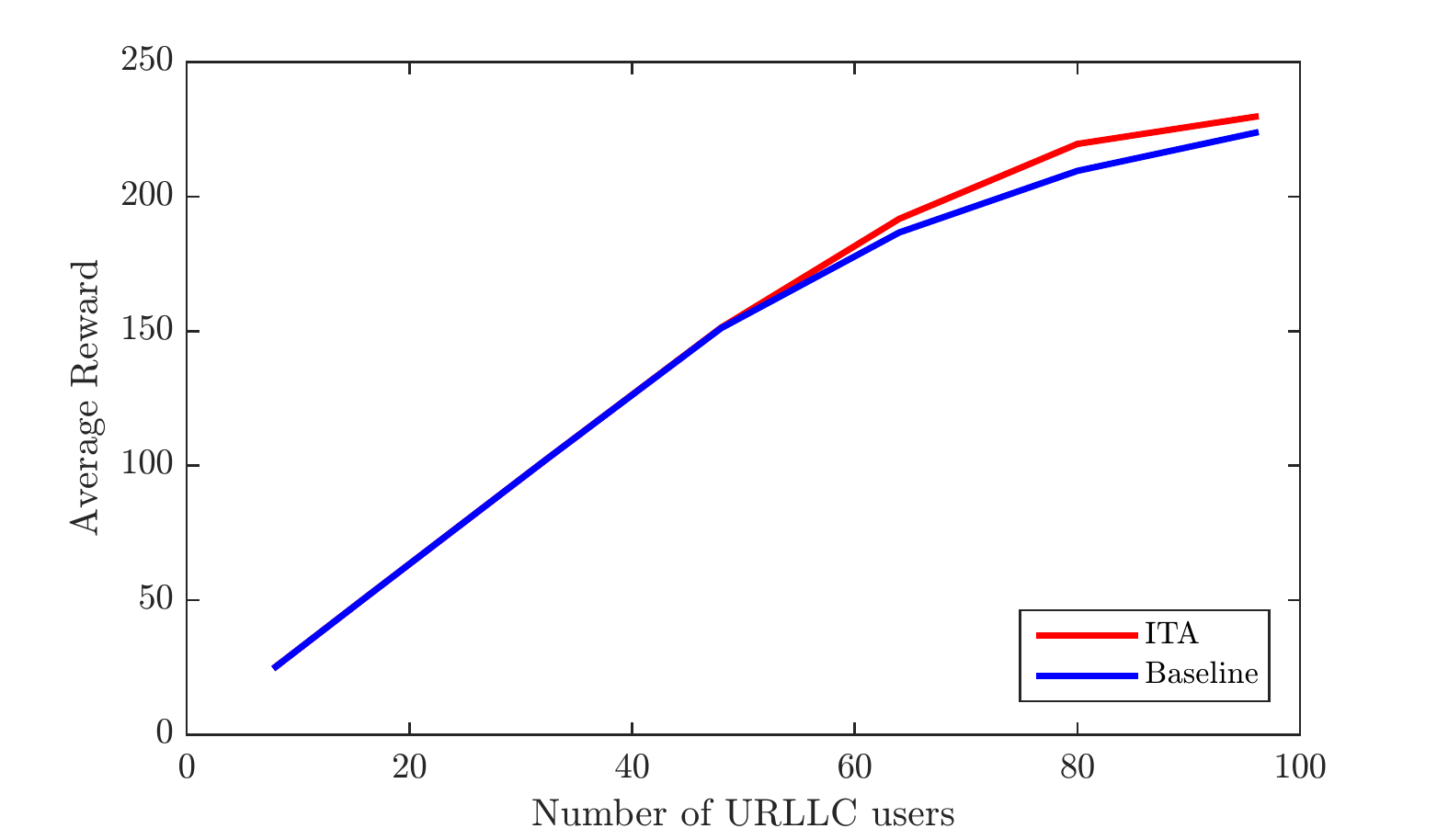}
		}
		\caption{Average reward gathered from admitting URLLC users with (a) low (b) medium and (c) high number of resources allocated to URLLC traffic. Each user has a packet size of $32${ Bytes } and requests $99.999\%$ reliability. The results shown are  averages over $5000$ trials with generated Rayleigh fading.} 
		\label{fig:randomRewards}
	\end{figure*}
	Finally, we examine a setting where users are placed randomly in a cell but the reward a user will bring if admitted is proportional to the logarithm of its mean SNR. The rationale here is that an operator may want to prioritize admitting URLLC users with good general channel conditions, since it is more sustainable to serve these users for their whole session duration. Results are presented in Fig. \ref{fig:randomRewards}. Same as before, ITA brings significant benefits over the baseline algorithm when the number of RBs becomes comparable or less than the number of users.

	\section{Conclusion}

	In this paper, we proved, via a reduction to the maximum weighted independent set problem, that joint admission control and scheduling of URLLC users with  time-frequency resource blocks is NP-hard even in a simplified problem, where resource blocks have binary SNRs and user $k$ requests at least $M_k$ resource blocks at state $1$ to satisfy her reliability constraints. However, for this binary SNR problem, checking if a set of users is feasible can be done in polynomial time via a proper LP. This implies that hardness comes mainly from the admission control part of the problem, and we proposed a greedy algorithm with a provable approximation ratio. Finally, for the problem with continuous SNRs, we proposed a heuristic that iteratively selects thresholds for the SNRs and iteratively solves a binary SNR problem. Our heuristic  outperforms the baseline in simulations. 

	\bibliographystyle{IEEEtran}
	\bibliography{urllc_ofdma_biblio}

\begin{thebibliography}{10}
\providecommand{\url}[1]{#1}
\csname url@samestyle\endcsname
\providecommand{\newblock}{\relax}
\providecommand{\bibinfo}[2]{#2}
\providecommand{\BIBentrySTDinterwordspacing}{\spaceskip=0pt\relax}
\providecommand{\BIBentryALTinterwordstretchfactor}{4}
\providecommand{\BIBentryALTinterwordspacing}{\spaceskip=\fontdimen2\font plus
\BIBentryALTinterwordstretchfactor\fontdimen3\font minus
  \fontdimen4\font\relax}
\providecommand{\BIBforeignlanguage}[2]{{%
\expandafter\ifx\csname l@#1\endcsname\relax
\typeout{** WARNING: IEEEtran.bst: No hyphenation pattern has been}%
\typeout{** loaded for the language `#1'. Using the pattern for}%
\typeout{** the default language instead.}%
\else
\language=\csname l@#1\endcsname
\fi
#2}}
\providecommand{\BIBdecl}{\relax}
\BIBdecl

\bibitem{3gppURLLC}
3GPP, ``{Study on Scenarios and Requirements for Next Generation Access
  Technologies (Release 15)},'' {3rd Generation Partnership Project (3GPP)},
  Technical Specification (TS) Group Radio Access Network 38.913, 06 2018,
  version 15.2.0.

\bibitem{Holfeld2016}
B.~Holfeld, D.~Wieruch, T.~Wirth, L.~Thiele, S.~A. Ashraf, J.~Huschke,
  I.~Aktas, and J.~Ansari, ``{Wireless Communication for Factory Automation: an
  opportunity for LTE and 5G systems},'' \emph{IEEE Commun. Mag.}, vol.~54,
  no.~6, pp. 36--43, Jun 2016.

\bibitem{Zhang2017}
H.~Zhang, N.~Liu, X.~Chu, K.~Long, A.~H. Aghvami, and V.~C.~M. Leung,
  ``{Network Slicing Based 5G and Future Mobile Networks: Mobility, Resource
  Management, and Challenges},'' \emph{IEEE Commun. Mag.}, vol.~55, no.~8, pp.
  138--145, 2017.

\bibitem{Campolo2017}
C.~Campolo, A.~Molinaro, A.~Iera, and F.~Menichella, ``{5G Network Slicing for
  Vehicle-to-Everything Services},'' \emph{IEEE Wireless Commun. Mag.},
  vol.~24, no.~6, pp. 38--45, Dec 2017.

\bibitem{Nielsen2018}
J.~J. Nielsen, R.~Liu, and P.~Popovski, ``{Ultra-Reliable Low Latency
  Communication Using Interface Diversity},'' \emph{IEEE Trans. Commun.},
  vol.~66, no.~3, pp. 1322--1334, Mar 2018.

\bibitem{Kotaba2018}
R.~Kotaba, C.~N. Manchón, T.~Balercia, and P.~Popovski, ``Uplink transmissions
  in {URLLC} systems with shared diversity resources,'' \emph{IEEE Wireless
  Commun. Lett.}, 2018.

\bibitem{Rao2018}
J.~Rao and S.~Vrzic, ``{Packet Duplication for URLLC in 5G: Architectural
  Enhancements and Performance Analysis},'' \emph{IEEE Netw.}, vol.~32, no.~2,
  pp. 32--40, Mar 2018.

\bibitem{Durisi2016}
G.~Durisi, T.~Koch, and P.~Popovski, ``{Toward Massive, Ultra-reliable, and
  Low-Latency Wireless Communication With Short Packets},'' \emph{Proc. IEEE},
  vol. 104, no.~9, pp. 1711--1726, Sept 2016.

\bibitem{Huang2009}
J.~Huang, V.~G. Subramanian, R.~Agrawal, and R.~A. Berry, ``{Downlink
  scheduling and resource allocation for OFDM systems},'' \emph{IEEE Trans.
  Wireless Commun.}, vol.~8, no.~1, pp. 288--296, Jan 2009.

\bibitem{She2017}
C.~She, C.~Yang, and T.~Q.~S. Quek, ``{Radio Resource Management for
  Ultra-Reliable and Low-Latency Communications},'' \emph{IEEE Commun. Mag.},
  vol.~55, no.~6, 2017.

\bibitem{She2018}
------, ``{Joint Uplink and Downlink Resource Configuration for Ultra-Reliable
  and Low-Latency Communications},'' \emph{IEEE Trans. Commun.}, vol.~66,
  no.~5, pp. 2266--2280, May 2018.

\bibitem{Arnau2018}
J.~Arnau and M.~Kountouris, ``{Delay performance of MISO wireless
  communications},'' in \emph{WiOpt}, May 2018.

\bibitem{Destounis2018}
A.~Destounis, G.~S. Paschos, J.~Arnau, and M.~Kountouris, ``{Scheduling URLLC
  users with reliable latency guarantees},'' in \emph{WiOpt}, May 2018.

\bibitem{Sharma11}
M.~{Sharma} and X.~{Lin}, ``Ofdm downlink scheduling for delay-optimality:
  Many-channel many-source asymptotics with general arrival processes,'' in
  \emph{ITA Workshop}, 2011.

\bibitem{Bodas2012}
S.~Bodas, S.~Shakkottai, L.~Ying, and R.~Srikant, ``Low-complexity scheduling
  algorithms for multichannel downlink wireless networks,'' \emph{IEEE/ACM
  Trans. Netw.}, vol.~20, no.~5, pp. 1608--1621, Oct 2012.

\bibitem{Anand2018_puncturing}
A.~Anand, G.~de~Veciana, and S.~Shakkottai, ``{Joint Scheduling of URLLC and
  eMBB Traffic in 5G Wireless Networks},'' in \emph{IEEE INFOCOM}, 2018.

\bibitem{Anand2018_ofdma}
A.~Anand and G.~de~Veciana, ``{Resource Allocation and {HARQ} Optimization for
  {URLLC} Traffic in {5G} Wireless Networks},'' \emph{IEEE J. Sel. Areas
  Commun}, vol.~36, no.~11, Nov 2018.

\bibitem{Fountoulakis17_scalable}
E.~Fountoulakis, N.~Pappas, Q.~Liao, V.~Suryaprakash, and D.~Yuan, ``{An
  examination of the benefits of scalable TTI for heterogeneous traffic
  management in 5G networks},'' in \emph{RAWNET}, 2017.

\bibitem{3gpp5GNumerology}
3GPP, ``{NR; Physical channels and modulation (Release 15)},'' {3rd Generation
  Partnership Project (3GPP)}, Technical Specification (TS) Group Radio Access
  Network 38.211, 06 2018, version 15.2.0.

\bibitem{Polyanskiy10}
Y.~Polyanskiy, H.~V. Poor, and S.~Verdu, ``Channel coding rate in the finite
  blocklength regime,'' \emph{IEEE Trans. Inf. Theory}, vol.~56, no.~5, pp.
  2307--2359, May 2010.

\bibitem{Yang2014_multichannel}
W.~Yang, G.~Durisi, T.~Koch, and Y.~Polyanskiy, ``Quasi-static multiple-antenna
  fading channels at finite blocklength,'' \emph{IEEE Trans. Inf. Theory},
  vol.~60, no.~7, pp. 4232--4265, Jul. 2014.

\bibitem{Austrin11}
P.~Austrin, S.~Khot, and M.~Safra, ``Inapproximability of vertex cover and
  independent set in bounded degree graphs,'' \emph{Theory Comput.}, vol.~7,
  pp. 27--43, 2011.

\bibitem{Khot2005}
S.~Khot, ``On the unique games conjecture,'' in \emph{FOCS}, 2005.

\bibitem{Papadimitriou_book}
C.~H. Papadimitriou and K.~Steiglitz, \emph{Combinatorial Optimization:
  Algorithms and Complexity}.\hskip 1em plus 0.5em minus 0.4em\relax Mineola,
  New York: Dover Publications Inc., 1998.

\end{thebibliography}
\end{document}